\documentclass[]{interact}

\usepackage{epstopdf}
\usepackage{caption}

\theoremstyle{plain}

\theoremstyle{definition}

\theoremstyle{remark}

\usepackage[section]{placeins}
\makeatletter
\AtBeginDocument{%
  \expandafter\renewcommand\expandafter\subsection\expandafter{%
    \expandafter\@fb@secFB\subsection
  }%
}
\makeatother

\usepackage{amsfonts}
\usepackage{fancyhdr}
\usepackage{comment}
\usepackage{times}
\usepackage{amsmath}
\usepackage{changepage}
\usepackage{amssymb}
\usepackage{graphicx}%
\usepackage{enumitem}
\usepackage{multirow}
\usepackage[colorlinks=true,urlcolor=blue]{hyperref}
\usepackage{pdfpages}
\usepackage[ruled,vlined]{algorithm2e}
\usepackage{ifthen}
\usepackage{tabularx}
\usepackage{array, ltablex, multirow} 
\usepackage{todonotes}
\usepackage{listings} 
\usepackage{xcolor}

\graphicspath{{figs/}} 

\usepackage{bm}

\newcommand{\Z}{\mathbb{Z}}
\newcommand{\D}{\mathbb{D}}

\newcommand{\etal}{\textit{et al.}}

\newcommand{\bx}{\bm{x}}
\newcommand{\bd}{\bm{d}}
\newcommand{\by}{\bm{y}}

\newcommand{\be}{\bm{e}}
\newcommand{\bz}{\bm{z}}
\newcommand{\bB}{\bm{B}}

\begin{document}


\title{OpenEMS: an open-source Package for Two-Stage Stochastic and Robust Optimization for Ambulance Location and Routing with Applications to Austin-Travis County EMS Data}

\author{
\name{Joshua Ong\textsuperscript{a}\thanks{Correspondence: Joshua Ong Email: joshong@utexas.edu}, David Kulpanowski \textsuperscript{b}, Yangxinyu Xie \textsuperscript{c}, Evdokia Nikolova \textsuperscript{a}, Ngoc Mai Tran \textsuperscript{d}} 
\affil{\textsuperscript{a}Department of Electrical and Computer Engineering, University of Texas at Austin, TX 78712, USA.; \textsuperscript{b}Department of Emergency Medical Services, City of Austin; 
\textsuperscript{c}Department of Computer Science, University of Texas at Austin, TX 78712, USA.; 
\textsuperscript{d}Department of Mathematics, University of Texas at Austin, TX 78712, USA.}
}

\maketitle

\begin{abstract}
Emergency Medical Systems (EMS) provide crucial pre-hospital care and transportation. Faster EMS response time provides quicker pre-hospital care and thus increases survival rate. We reduce response time by providing optimal ambulance stationing and routing decisions by solving two stage stochastic and robust linear programs. Although operational research on ambulance systems is decades old, there is little open-source code and consistency in simulations. We begin to bridge this gap by publishing OpenEMS, in collaboration with the Austin-Travis County EMS (ATCEMS) in Texas, an end-to-end pipeline to optimize ambulance strategic decisions. It includes data handling, optimization, and a calibrated simulation. We hope this open source framework will foster future research with and for EMS. Finally, we provide a detailed case study on the city of Austin, Texas. We find that optimal stationing would increase response time by 88.02 seconds. Further, we design optimal strategies in the case where Austin EMS must permanently add or remove one ambulance from their fleet.
\end{abstract}


\section{Introduction}

The Emergency Medical Service (EMS) provides pre-hospital treatment and transportation. 
Receiving fast pre-hospital treatment is vital in scenarios such as cardiac arrest, stroke, and other severe trauma cases \cite{Mahama2018Dec}. It is generally agreed upon that these calls need pre-hospital treatment within 8-10 minutes \cite{Pons2002Jul, Courtemanche2019Sep, bertsimas2019robust, Lam_2016}. If a call is not answered with in this threshold, it is considered a shortfall. To help EMS services achieve the best response time the operational research community has widely studied how ambulance stationing and routing can optimize response times. \cite{bertsimas2019robust,BERALDI2004183,LIU201979, Morohosi2008ACS,BURKEY_2012,Nilsang_2018,SWALEHE2016199,mccormack_2015,Ingolfsson_2003,Inakawa_2010}.

Strategic decisions can also help EMS departments save money. White ~\etal \cite{White_2007} found health care spending has dramatically increased in the US compared to other countries. Further, Heightman \cite{Heightman2009Mar} approximates that it costs half a million dollars a year to staff one ambulance on a 24/7 basis. One way to combat increased spending is to use current EMS resources more efficiently. Optimization could help a city with a decreasing budget maintain performance while removing an ambulance from its operation.      

The operational research community commonly constructs the ambulance problem into a two-stage optimization problem \cite{bertsimas2019robust,BERALDI2004183,LIU201979}. 
The first stage is the ambulance location problem (ALP). 
This answers the question: {\em where should we station our ambulances around the city to achieve an optimal coverage?}
Ambulances can be stationed around the city in locations such as hospitals, fire stations, ect. 
In the ALP we want to maximize the coverage of the ambulances to respond to calls, using historic call data as empirical probability distributions.

\subsection{Our Contributions}

There have been many case studies on ambulance stationing and routing \cite{Morohosi2008ACS,BURKEY_2012,Nilsang_2018,SWALEHE2016199,mccormack_2015,Ingolfsson_2003,Inakawa_2010}, but the methodologies regarding discrete-event simulations vary widely. Some are theoretical and include no simulations, some use private softwares, while only a few provide open-source tools \cite{bertsimas2019robust}. In this work, we develop OpenEMS, a modular software with open-source code consisting of four well-separated parts: data handling, optimization, numerical computations of discrete time events, and visualization. This allows future research that innovates on one or several components to be directly compared with previously published work. We hope that this increased transparency would facilitate the translation from research results to field implementations by EMS departments. Our main contributions are as follows:

\begin{enumerate}
\setlength\itemsep{.1em}
    \item A software pipeline at \url{https://github.com/joshua-ong/AmbulanceDeployment} that can be applied to any city. The software includes: discretizing a set area into a grid, calibration, validation, optimization using state-of-the-art (SOTA) methods from Bertsimas~\etal \cite{bertsimas2019robust}, simulation, and visualizing improvements compared to historical performance. 
    \item  A case study of Austin EMS, including 246,809 thousand calls within 1039 square miles, demonstrating the impact of optimal decision making.
\end{enumerate}

The rest of the paper is organized as follows. Section ~\ref{sec:Past} gives a brief overview of ambulance dispatch research. Next, we include the necessary theory of the optimization models in Section ~\ref{sec:Optimization}. We detail the simulation model in Section ~\ref{sec:Simulation}. Next, we have a case study of the city of Austin in section ~\ref{sec:Case}. Next, we include a socio-economic of Austin in section ~\ref{sec:socio_economic}. Next, there is a user code sample in section ~\ref{sec:UserGuide}. We conclude in Section ~\ref{sec:Conclusion}.

\section{Past Work}\label{sec:Past}

There exist many case studies across various countries showing that EMS systems' heuristics are not optimal. Morohosi \cite{Morohosi2008ACS}, Eaton~\etal \cite{Eaton_1985}, and Burkey ~\etal \cite{BURKEY_2012} use popular coverage linear programs to study Tokyo, Japan; Austin, Texas; and 4 states in the southern USA respectively. Eaton~\etal \cite{Eaton_1985}, and Burkey ~\etal \cite{BURKEY_2012} also use coverage to examine social equality. Nilsang ~\etal \cite{Nilsang_2018} included a case study in Bangkok, Thailand where they utilize social media to design a covering model. They found they could reduce the number of ambulance locations by 81.6 \% without reducing coverage. Swalehe ~\etal \cite{SWALEHE2016199} included a case study in Eskisehir Province, Turkey using GIS (Geogrpahic Information System) methods and found they could reduce the response time from 10 minutes to 5 minutes for 77.6 \% of ambulance demand areas. McCormack ~\etal \cite{mccormack_2015} included a case study in London, England using genetic algorithms and testing the algorithm on a calibrated and validated simulation model. Doener ~\etal \cite{Doerner} includes a broader case study of Austria including coverage linear programs for ambulance systems but also other related areas like inventory routing and disaster relief. Nonetheless, none of these papers include open-source code.

These case studies can also answer important hypothetical questions. McCormack ~\etal \cite{mccormack_2015} used simulations to find how performance would decrease if there were resource gaps that day. Ingolfsson ~\etal \cite{Ingolfsson_2003} tested the policy of a 'single start station', where all ambulances start at the same station using the  ProModel software \cite{Price2000}. Although the policy of having all ambulances start at the same station seems inefficient, it has the benefits of having pooled resources in one location. Inakawa ~\etal \cite{Inakawa_2010}  used simulation on the city of Seto, Japan to find that adding a new location was twice as effective as adding two ambulance stations. Aboueljinane ~\etal \cite{Aboueljinane2012R} used simulation in Val-de-Marne department of France using the ARENA software \cite{Markovitch1995}. In it they recommended repositioning teams to increase coverage and found that it improved response time by 3.5$\%$.  

Implementing novel algorithms into an EMS department has proven to be difficult. For example, in 1992, the London Ambulance Service (LAS) tried to implement a computerized system including a Computerized Dispatch System (CAD) (rather than human operators), and automatic vehicle stationing \cite{Fitzgerald2005Sep}. According to Adamu ~\etal \cite{Fitzgerald2005Sep} the software had a $\pounds$ 7.5 million budget, but only lasted 9 days before it was closed due to failure and increased mortality. The software was abandoned after 9 days as implementations did not have time to learn and fail, because there are human lives at stake. It further shows the significance of testing novel algorithms on confident simulations. 

A notable success began with Mason and Henderson's \cite{Henderson2005} simulation model BartSim, which later matured into a company \cite{Mason_2013} called Optima. Optima currently has case studies in New Zealand, Australia, Denmark, United Kingdoms, Canada, and the United States. However, Optima is a commercial software designed for EMS operators. In contrast, OpenEMS is open-source and designed to support researchers making the latest advances in the ambulance allocation and routing problem with data validations while incurring minimal technical setup and no charge. 

Some of the earliest work began when ReVelle~\etal \cite{ReVelle_1989} proposed the Maximum Available Location Problem (MALP), inspired by the more general Maximum Coverage Problem (MCP). In MALP they optimize the location of stations for ambulances to maximize the coverage, where a caller is covered if there is an ambulance that can reach the call within a fixed time. This is called the ambulance location problem (ALP). Later Batta~\etal \cite{Batta1989} introduced Maximum Expected Covering Location Problem (MEXCLP), which used an ambulance's "busy fraction" as a Bernoulli random variable to approximate the expected coverage. Other classical models include the Double Standard Model (DSM) by Gendreau ~\etal \cite{GENDREAU199775}. These remain the important benchmarks in many case studies \cite{bertsimas2019robust,Morohosi2008ACS,Eaton_1985,BURKEY_2012}.

Further research from the operational research community has since been applied to solving the ambulance problems. Beraldi~\etal \cite{BERALDI2004183} solved the ALP with chance constraints to model routing ambulances to an unknown demand. For more information on two stage stochastic formulation, we refer the reader to \cite{ahmed} and \cite{Birge_2011}. A two-stage robust formulation was first implemented by Bertsimas~\etal \cite{bertsimas2019robust} who applied a data-driven model to generate scenarios. A two-stage distributional robust formulation was implemented by Liu~\etal \cite{LIU201979}.

Other studies integrate a tool called the approximate dynamic programming (ADP). Some of the first papers on applying ADP to the ambulance problem were from Maxwell ~\etal \cite{Maxwell_2010} and Schmid ~\etal \cite{SCHMID2012}. Later Jagtenberg ~\etal \cite{Jagtenberg_2017Dec} used ADP to show that the closest ambulance routing policy is not necessarily optimal. For a more comprehensive overview, we refer readers to the review by Aboueljinane~\etal \cite{ABOUELJINANE_2013} and Tassone ~\etal \cite{tassone2020comprehensive}.

\section{Optimization}\label{sec:Optimization}

In this section, we lay out the theoretical formulation of two-stage stochastic and robust formulations appeared in Bertsimas ~\etal \cite{bertsimas2019robust} to solve ALP and ARP. We adopt their notation in this paper. In a two-stage optimization problem, there is the first stage $\emph{here-and-now}$ decision which must be made under uncertainty and the $\emph{wait-and-see}$ variable that happens afterwards.

In our context, at the first stage we must select how many ambulances are stationed at each location. Given the set of ambulance stations $I$; let $x_i$ be the number of ambulances at location $i \in I$. This is the first stage decision, as we must station the ambulances before we know the realization of the demand. In the second stage the calls are revealed and we must determine the optimal routing of stationed ambulances to these calls. Let $J$ be the set of regions, then each call can be assigned to a region $j \in J$. Let $y_{ij}$ be the routing of ambulance stationed at $i$ to region $j$. Let $\tau$ be the longest time it takes for an ambulance to serve a call and return to their station. Then we can group calls within periods of $\tau$ where we assume an ambulance can serve one call in each period. A complete table of notation is included in table \ref{tab:notation}.

\begin{table}[ht]
    \centering
    \begin{tabular}{|l|l|} 
    \hline
      \textbf{Name} & \textbf{Description}\\
      \hline
      $I$ & Set of ambulance locations  \\
      $J$ & Set of demand regions \\
      $E$ & Set of feasible edges from $I$ to $J$\\
      $M$ & Set of observed ambulance calls \\
      $n$ & Total number of ambulances  \\
      $\tau$ & Total time (minutes) for an ambulance to serve a call and be available again  \\
      $y_{ij}$ & Number of ambulances routed from location $i$ to region $j$   \\
      $x_i$ & Number of ambulances at location $i$  \\
      $d_j$ & Number of ambulance calls from region $j$ \\
      $\bm{B}_I$ & Incidence matrix representing incoming edges \\
      $\bm{B}_J$ & Incidence matrix representing outgoing edges \\\hline
    \end{tabular}
    \caption{Summary of notation for ambulance location and dispatch.}
    \label{tab:notation}
\end{table}
  
\subsection{Stochastic Formulation}

The stochastic formulation objective is to minimize the expectation of shortfalls. To solve this problem we use Monte-Carlo sampling to construct a discrete distribution $\hat{\mathbb{P}}(\hat \bd)$ where we assume each sample of calls is equally likely. The stochastic formulation is written as:

\begin{subequations}
\begin{equation}
\text{minimize}  \quad \frac{1}{M} \sum_{m=1}^M \bz^m
\end{equation} 
\label{eq:stocha}  
\begin{equation}
\text{subject to} \quad \displaystyle \be^{\top}\bx \leq n 
\label{eq:stochb} 
\end{equation}
\begin{equation}
\quad \bx - \bB_I \by^{m} \geq 0 \>\> \forall m\in [M] 
\label{eq:stochc} 
\end{equation}
\begin{equation}
\quad \bz^m \geq \bd^m - \bB_J \by^m \>\> \forall m\in [M]
\label{eq:stochd} 
\end{equation}
\begin{equation}
\quad \bx \in \Z_+^{I}, \by^m \in \Z_+^{E}, \bz^{m} \in \Z_+^J \>\> \forall m\in [M]
\label{eq:stoche} 
\end{equation}
\label{eq:stoch}
\end{subequations}

The objective \ref{eq:stocha} contains a slack variable that is bounded from below later. Constraint \ref{eq:stochb} limits the number of ambulances stationed to be less than or equal to the total number of ambulances. Constraint \ref{eq:stochc} iterates through each Monte-Carlo sample $(m \in M)$, for each sample we limit a stationed ambulance to only be sent once per time period, $\tau$. Constraint \ref{eq:stochd} makes the slack variable greater than or equal to the shortfalls, which is equal to calls minus calls answered. (Note this can be thought of as a maxflow problem of a bipartite graph, where $x$ and $d$ are the vertices of the bipartite graph and $y$ are the edge values.)

\subsection{Robust Formulation}

In the robust optimization formulation, we want to minimize the worst-case scenarios. To do this we first model demand as a Poisson distribution, $\gamma$. (Single is the demand of a single grid cell. Local is the demand of adjacent grid cells. Regional is the demand within 10 minutes of a grid cell. Global is the demand of all of Austin.) We then construct the uncertainty set $\D (\alpha)$ as the $\alpha$ Value-at-Risk of these Poisson distributions. Finally, we take the worst scenarios from and minimize over them in the robust formulation like:	

\begin{subequations}
\begin{equation}
\text{maximize}_{\bd \in \D (\alpha)} \text{minimize}  \quad \frac{1}{M} \sum_{m=1}^M \bd^m - \bB_J \by^m 
\label{eq:robusta}  
\end{equation} 
\begin{equation}
\text{subject to} \quad \displaystyle \be^{\top}\bx \leq n 
\label{eq:robustb} 
\end{equation}
\begin{equation}
\quad \bx - \bB_I \by^{m} \geq 0 \>\> \forall m\in [M] 
\label{eq:robustc} 
\end{equation}
\begin{equation}
\quad \bz^m \geq \bd^m - \bB_J \by^m \>\> \forall m\in [M] 
\label{eq:robustd} 
\end{equation}
\begin{equation}
\quad \bx \in \Z_+^{I}, \by^m \in \Z_+^{E}, \bz^{m} \in \Z_+^J \>\> \forall m\in [M]
\label{eq:robuste} 
\end{equation}
\label{eq:robust}
\end{subequations}

where: 

\begin{multline}
  \D(\alpha) = \{\bd\in\Z^{J}_+ : d_j \le VaR(\gamma_j^{\text{single}}(\alpha)) \>\> \forall j \in J,  VaR(\gamma_j^{\text{local}}(\alpha)) \>\> \forall j \in J, \\ 
  d_j \le VaR(\gamma_j^{\text{regional}}(\alpha)) \>\>\forall i \in I, d_j \le VaR(\gamma_j^{\text{global}}(\alpha)\}
\end{multline}

The formulation is the same as the stochastic formulation except for the objective. For more information we refer the reader to the original formulation by Bertsimas et al. \cite{bertsimas2019robust} on how to dualize \ref{eq:robust} and make it tractable with the column-constraint method. Zeng and Zhao \cite{ZENG2013457}  first designed the column-constrained method for two-stage robust optimization. The column-constraint method has the same idea as the Benders-style but with faster heuristic implementation. 

\section{Simulation Model}\label{sec:Simulation}

In this section, we describe the different components of the simulation model in the general setting. This includes the steps to replicate this on a different case study. Notably, we measure the confidence of the simulation response time so that an EMS department can understand the  possible variance in real-world applications. First, we explain the state space of a general Emergency Medical System (EMS) and the correspondence of these events to the simulation. Next,we explain how the simulation represents the real world by discretizing it into a grid. We then verify the model by simulating historical actions and comparing them with what happened to validate that the model works in normal conditions. Finally, we explain the process of tuning the linear program parameters for best results.  

\subsection{The state space for answering an emergency call} 

\begin{figure}[ht]
  \includegraphics[width=1.0 \textwidth]{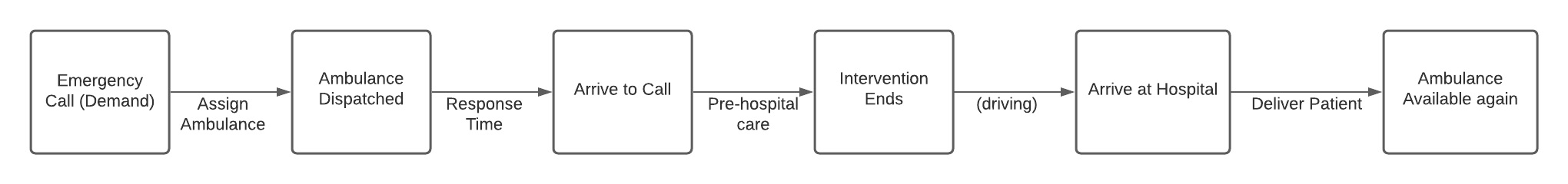}
  \caption{A general state space of the EMS procedure for answering an emergency call.}
  \label{fig:statespace}
\end{figure}

A general framework of the EMS procedure for answering an emergency call is detailed in figure \ref{fig:statespace}. The first state includes from the time the call was made to when it was assigned to an ambulance, where an operator must collect enough information like the seriousness of the injury and location. The time the ambulance is en route to when it arrives at the call is labeled the response time. This is arguably the most important metric for survival rate of the patient. From the time the ambulance arrived to when it departs the scene captures the time it took to administer pre-hospital treatment and load the patient. The time departed from the scene to the hospital is drive time. The ambulance is now available to serve another call, if there are no new calls, it will return to its station.

Our simulation model includes the same steps as the state space just described. Each step is an event that is put into a priority queue with respect to time. An overview of the code is included in pseudo-code \ref{fig:algorithm}. In the beginning, all calls are input into the priority queue at their respective times. The calls then propagate the other events following the state space. We do not include the setup time, because it is only a fraction of the whole ambulance travel time and is not impacted by our strategic decision. The travel times are determined using Open Street Map, which was introduced by Haklay and Weber \cite{Haklay2008OpenStreetMapUS}. The intervention time is modeled as a lognormal distribution. They have been shown to effectively fit ambulance processes like in Lam ~\etal \cite{Lam2016} and McLay and Mayorga \cite{McLay_2013}.

\begin{algorithm}[ht]
\SetAlgoLined
 priorityQueue.enque(allEmergencyCalls) \\
 \While{priorityQueue not empty}{
  event = priorityQueue.pop() \\
  event.handle() \\
  \If{event == newCall}{
    priorityQueue.push(callEnroute)
  }  
  \If{event == callEnroute}{
    priorityQueue.push(callArriveScene)  
  }
  \If{event == callArriveScene}{
    priorityQueue.push(callDepartScene)  
  }
  \If{event == callDepartScene}{
    priorityQueue.push(callArriveHospital)  
  }
  \If{event == callArriveHospital}{
    priorityQueue.push(AmbulanceAvailable)  
  }
 }
 \caption{An overview of the simulation model as pseudo code.}
 \label{fig:algorithm}
\end{algorithm}

\subsection{Grid construction and travel time}\label{sec:gridConstruction}

A grid construction is necessary for both solving the linear programs and modeling the simulation. The main idea is to discretize space by representing the continuous space with a grid, like in figure \ref{fig:grid_plot}. If an event happens somewhere within a grid cell, we assume it happens at the center of the grid cell. Discretizing the ambulance space is common in both linear programming theory \cite{bertsimas2019robust} as well as other simulation models \cite{mccormack_2015}. 

To construct the grid, first we inscribe the case study region into a rectangle. Then we split the grid into uniformly sized rectangle members. Note there is a tradeoff between how much resolution (number of grid cells) to include and how much computation is required. If the grid is finer it has more accurate travel times but longer computations and vice versa. Once the grid is established, any call within a rectangle member will be considered to be at the center. Then the travel time between any two points becomes the distance of the members of the grid cells they belong to. 

We compute travel times with Open Street Map (OSM). In general, we find computed travel times are slower than the reported travel time. Brown ~\etal \cite{BROWN2000} found that ambulances using lights and sirens traveled on average 105 seconds faster than ambulances without. This is to say that on average the special privileges of an ambulance allow it to go faster by bypassing traffic. Thus we recommend adjusting grid times to be more accurate. Regression will depend on the data available and other apriori knowledge, but can be relatively simple including temporal variables like day-of-week and hour or spatial variables like the number of grid cells away the path is.


\subsection{Verification}\label{sec:verification}

The verification step builds confidence that the model reflects the real world. It evaluates how well the calibration step compares to real world measurements. The verification model inputs historical actions and compares the simulated times with the reported times. Because there is a lot of randomness in the drive time to do with paths and traffic, we compare the overall trends for $n$ = 1000 calls or about two weeks. We then look at the mean response time error on average. To measure confidence, we then invoke Multiple Replication Procedure (MRP) on multiple batches of $n$-calls and measure the variance among batches. 

\subsection{Alpha Value}

As suggested by Bertsimas ~\etal \cite{bertsimas2019robust}, the $\alpha$ hyperparameter is determined experimentally for the robust linear program. This is done using cross-validation. In step one we choose a list of alpha values, for example $[0.1,.01,.001]$. In step two we split the dataset into a $k$-fold of training and testing sets with an $80/20$ split. We optimize the robust optimization on the training set and simulate it with our model on the testing set. We then remove any alpha values that are saturated (every station includes exactly one ambulance) and choose the alpha value which gives the best performance. 

\section{Case Study}\label{sec:Case}

\subsection{Motivation on Austin EMS}

In this section, we provide a case study of Austin, Texas, as an application of our open-source OpenEMS. Austin-Travis County EMS (ATCEMS) \footnote{https://www.austintexas.gov/content/ems-austin-travis-county} has 37 full-time ambulances that on average serve 338 calls per day. Data in this study includes 246,809 calls from the 2019-2020 period. ATCEMS serves 2.2 million residents within a 1,039 square mile space. Since Austin is one of the biggest cities in the United States, it is an interesting and representative subject to study.

\subsection{Data and statistics} \label{sec:data}

\begin{figure}[ht]
\centering
\includegraphics[width=1.0 \textwidth]{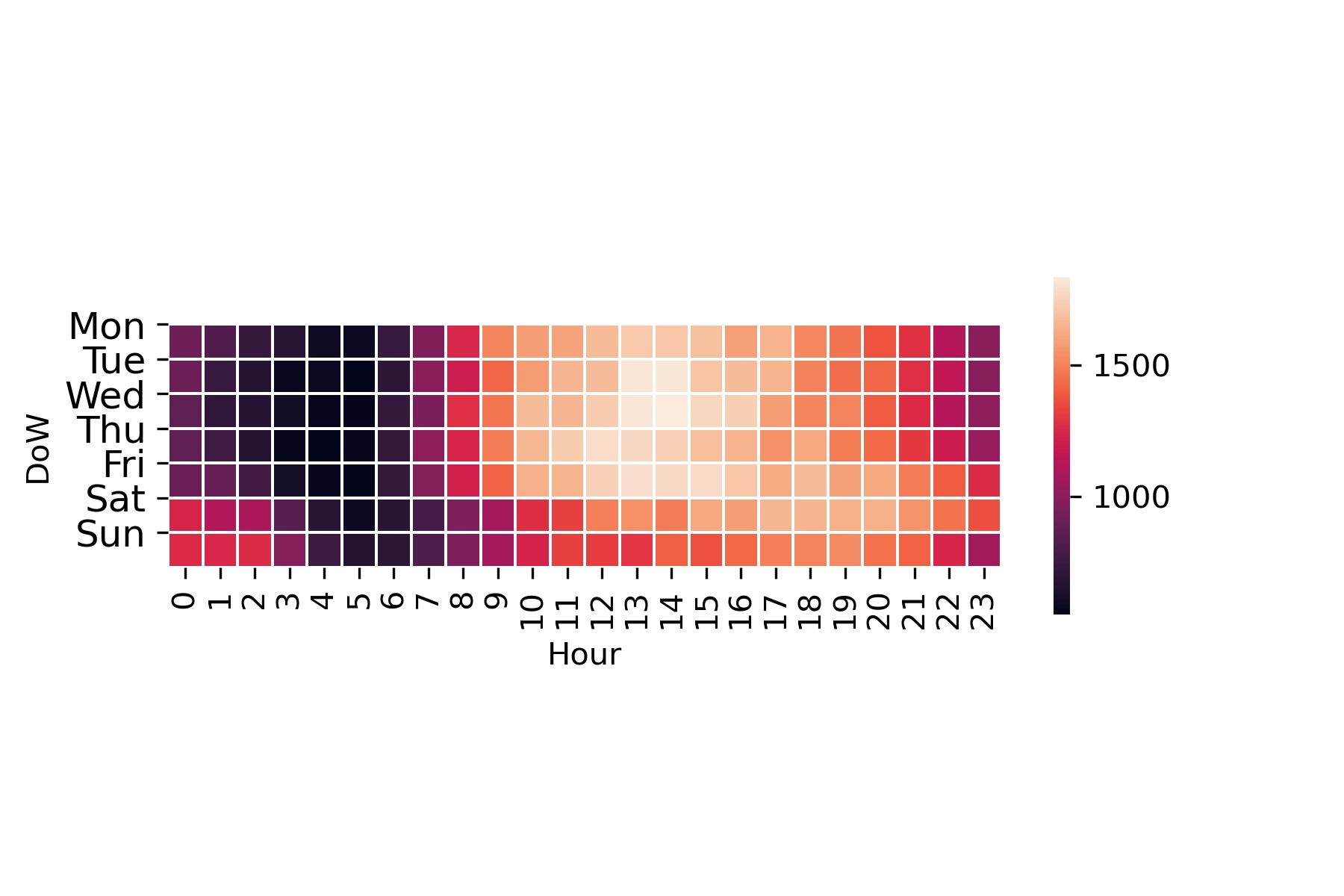}
\caption{A heat map showing the frequency of calls with respect to day of week and call hour.}
\label{fig:temporal_distribution}
\end{figure}

\begin{figure}[ht]
\centering
\includegraphics[width=1.0 \textwidth]{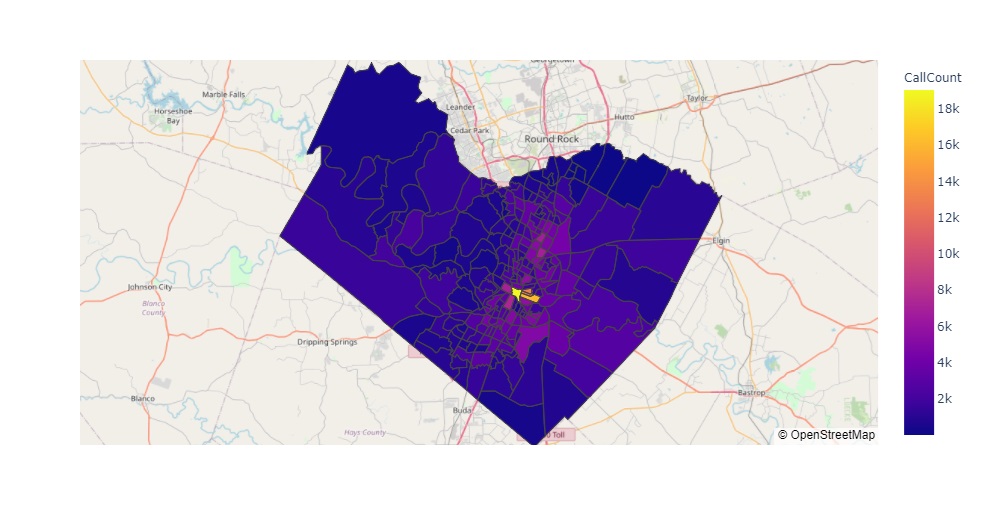}
\caption{A heat map showing the frequency of calls with respect to the spatial distribution.}
\label{fig:geo_distribution}
\end{figure}

Here we give an outline of what publicly available data exists. The main data entries included were the date of event, longitude, and latitude of call, and reported ambulance response time. (There also exists other entries such as reported ambulance assign time, reported ambulance on-site time, reported ambulance hospital time, ect.) We then graphed all calls and constructed the temporal distribution in figure \ref{fig:temporal_distribution}. From this figure, we noticed that the most consistently high demand calls are from 8:00-20:00 and from Monday-Friday. We call these times peak hours because they have more consistently high call demand than the offpeak hours. Further, we remove Saturday and Sunday because the times of high demand are less consistent. For the following results, we limit our experiments to these peak hours as they are the times which Austin EMS system is most stressed. Offpeak hours tend to be slower and easier to manage. We also plotted the data geospatially in \ref{fig:geo_distribution}. Like most cities, a majority of calls happen in the downtown region. Austin also includes a sprawling suburb area that has relatively fewer calls but takes a long time to drive between. This makes Austin hard to solve using heuristic planning to handle both the high-demand area of downtown while still being able to handle the occasional suburban calls.

\subsection{Calibration via travel time adjustment} \label{sec: case_study_callibration}

The calibration step is meant to calibrate the grid time to the reported travel time. To do this we tried several regression models to fit our grid time with the reported travel time. In general, we find that there is a low correlation because while grid time is deterministic, the reported travel time is dependent on many unknown factors such as traffic. We tried including several other variables like time of day and the spatial distance from two points but we found there were marginal benefits.

\begin{figure}[ht]
\centering
\includegraphics[width=.33 \textwidth]{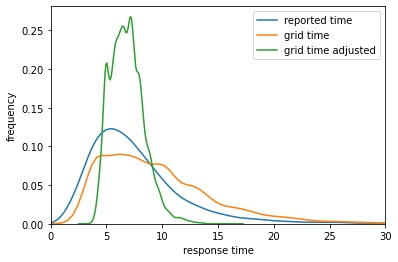}
\caption{An illustration of how the regression transforms the grid time.}
\label{fig:regression results}
\end{figure}

For this case study, we found that trimming the top and bottom 1 percent of data was necessary to avoid outlier reported travel times like 0 minutes (which is impossible) or calls that took more than an hour. Then we fitted the log of grid time to the log of reported time. The $r^2$ value was $.1472$. One possible reason is that the variance is mostly not dependent on the average grid time, but on unobserved factors such as traffic. We computed the adjusted grid times and compared them to the reported grid times in figure \ref{fig:regression results}. From this we observe that the adjusted times are squeezed with shorter tailends than before. From the validation we argue this produces a fair mean response time, but loses some resolution of extreme events as well as skewing times around 10 minutes which makes it harder to calculate shortfalls.  

\subsection{Verification} \label{sec: case_study_verification}

\begin{figure}[ht]
\centering
\includegraphics[width=.33 \textwidth]{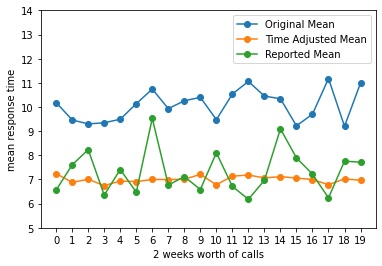}
\caption{This is a plot of 20 batches where each batch is one data point.}
\label{fig:verification}
\end{figure}

The results of the verification are shown in figure $\ref{fig:verification}$. Each data point represents the mean response time of 1000 calls or approximately 2 weeks' worth of data. Before the grid time had an error of $-186.23 \pm 70.13$ seconds, while with the adjusted grid time the error becomes $2.41 \pm 35.31$. From this, we see the calibration reduces the error. Thus we could expect the simulation mean response time to be close to the actual events within a quantified confidence.

\subsection{Alpha Value}

Next, we determine the best $\alpha$ value experimentally with a 3-fold cross-validation for the robust optimization. Results are in the appendix \ref{tab:cross_validation}. The set of optimal stationing was then tested on the same hold-out set. Interestingly, a majority of the optimizations have the same optimal value. This is the case except for when $\alpha$ becomes too big (.1) where it may not predict as many scenarios and when $\alpha$ becomes too small (.0001) where it may predict for too extreme scenarios. From this cross-validation we choose $\alpha = .01$ for the following experiments.

\subsection{Computational Setup}

The code was primarily written in Julia \cite{bezanson2015julia} with auxiliary scripts in Python. “Julia For Mathematical Programming” (JuMP) \cite{lubin_2015} was used to interface Gurobi 9.0 with Julia. 
The computations are done on a standard computer. The station has 16 GB of RAM and a 2.5 GHz processor.

\subsection{Experimental Setup}

To compare with real data we use 40 ambulances, similar to Austin EMS during the day. Simulations are run for 1000 calls. Then 12 batches of 1000 call simulations are run to establish confidence using Multiple Replication Procedure (MRP). (Note if a simulation is run on too few samples it is biased to perform better because all of the ambulances are available in the beginning. Then it will reach an equilibrium. If shortfalls become too numerous in a certain region, they can plug the system and give a bias towards worse performance.) The time it takes to serve is based on a lognormal distribution with a mean 3.65 and variance 0.3. Calls are filtered to peak hours to be more homogenous (see subsection \ref{sec:data}). In this simulation, the closest ambulance is always sent to the call. All parameters are summarized in table \ref{tab:Parameters}.

\begin{table}[ht]
    \centering
    \begin{tabular}{|l|l|}\hline
    Variable Name & Value \\\hline\hline
    number of ambulances & 40 \\
    number of calls & 1000 \\
    number of batches & 12\\
    serve time distribution & log-normal($\mu = $ 3.65, $\sigma = $ 0.3) \\
    $\alpha$ & .01\\
    \hline
    \end{tabular}
    \caption{Summary of experimental parameters.}
    \label{tab:Parameters}
\end{table}

\subsection{Numerical results} \label{sec: case_study_results}

Results from the simulation can be summarized in table \ref{tab:MRP}. We found the stochastic optimization performed better than robust optimization. Although, the robust optimization sometimes outperformed the stochastic optimization when there were more ambulances. This may be because robust optimization is more conservative, waiting for an anticipated scenario that does not happen in these batches. Both results perform better than the reported call time though. Namely, the stochastic optimization is expected to have a 88.02 second faster mean response time than heuristic decision making. 

\begin{figure}[ht]
    \centering
    \includegraphics[width=0.48\textwidth]{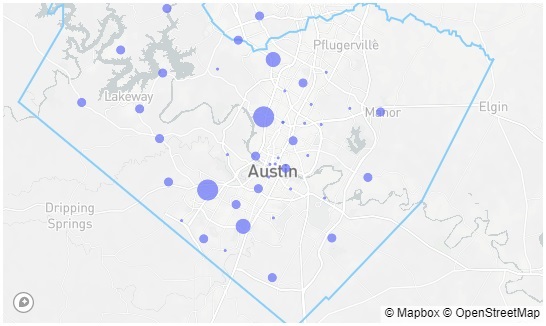}
    \includegraphics[width=0.48\textwidth]{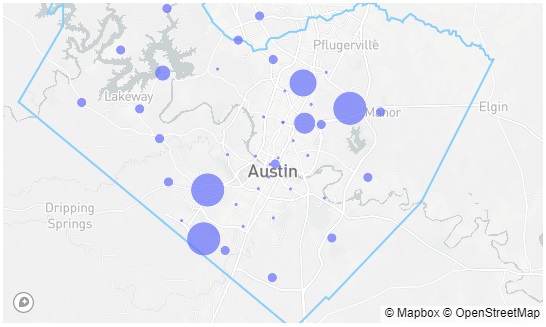}
    \caption{Ambulance location schemes for the Stochastic model (left) and Robust ($\alpha$ = .01) mode (right) for $n=40$. 
    The purple circle indicates there is at least one ambulance located in the area. 
    The radius of the circle is proportional to the number of ambulances in the area.}
    \label{fig:map_amb_40}
\end{figure}

The ambulance stationing was visualized in figure \ref{fig:map_amb_40}. The purple circles are proportional to how many ambulances should be stationed at each location. We note that the robust optimization will concentrate more ambulances in one station. We believe this is to handle some anticipated extreme scenarios that the robust optimization generated. Implementation of these policies is easy, as they only need to change the ambulance stationings to the optimal ones we calculated.

\begin{table}[ht]
    \centering
    \begin{tabular}{|l|l|l|l|}\hline
    Batch Number & Reported MRT & Stochastic Mean Response Time & Robust MRT \\\hline\hline
    Batch 1 & 8.021 & 6.145 & 6.706 \\
    Batch 2 & 8.418 & 6.22 & 6.747 \\
    Batch 3 & 7.306 & 6.224 & 6.76 \\
    Batch 4 & 8.539 & 6.239 & 6.81 \\
    Batch 5 & 7.385 & 6.185 & 6.769 \\
    Batch 6 & 7.799 & 6.2 & 6.831 \\
    Batch 7 & 7.512 & 6.24 & 6.787 \\
    Batch 8 & 7.276 & 6.134 & 6.71 \\
    Batch 9 & 7.891 & 6.264 & 6.856 \\
    Batch 10 & 7.064 & 6.231 & 6.893 \\
    Batch 11 & 7.423 & 6.229 & 6.704 \\
    Batch 12 & 7.495 & 6.21 & 6.828 \\
    Overall & 7.677 +/- 0.443 & 6.210 +/- 0.037 & 6.783 +/-  0.059\\
    \hline
    \end{tabular}
    \caption{The results of MRP batches on the simulation.}
    \label{tab:MRP}
\end{table}

\subsection{Adding or removing ambulances.}

Simulations can also help a city understand what to do when they have to permanently add or remove an ambulance from their fleet. An ambulance may be added if a city's population is growing and the department has new funds to employ a new ambulance engendering the question, where should they put it? The opposite scenario is when a department budget is cut and they must remove an ambulance, which one can they possibly remove to have the least impact? Thus optimization can help a city with a small budget maintain performance while removing an ambulance from operation.  

\begin{figure}[ht]
\centering
\includegraphics[width=.5 \textwidth]{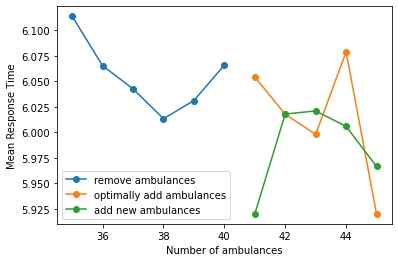}
\caption{The effects of adding or removing ambulances with Austin EMS.}
\label{fig:hypothetical}
\end{figure}

We conduct the optimal stationing as the number of ambulances changes in figure \ref{fig:hypothetical}. As can be seen, both the adding and the removal of ambulances only cause minor changes. This is good in the sense that if an ambulance must be removed performance does not decrease that much. But disappointingly adding ambulances also does not increase performance. These differences we found may be magnified though, as the grid time adjustment we saw earlier tends to "squeeze" the times together. 

\section{Understanding socio-economic factors}\label{sec:socio_economic}

To test whether socio-economic factors play a role in influencing the EMS response time across different neighborhoods across Austin, we applied the following protocol:
    \\ \\
    \textbf{Dataset} \quad
    For each census tract, we calculated the mean reported travel time of all incidents located within the census tract as the dependent variable. As for the independent variables, we included the minimum grid time and the average grid time within each census tract. This was calculated in two steps. First we calculated the minimum and average grid times for each grid cell. Then we took the weighted average (based on the call frequency) of grid cells within each census tract to represent them. Furthermore, we included 19 socio-economic independent variables extracted from the 2018 CDC Social Vulnerability Index dataset \cite{CDCSVI2018}. The variables included in this study are listed in Table \ref{tab:SVI list} in the appendix. \\ \\
    
    
    \noindent \textbf{5-fold Cross Validation} \quad
    We applied $5$-fold cross-validation. For each fold, we included $80\%$ as the training set and the remaining $20\%$ as the test set. Before training predictive models, we normalized each variable of the training set by subtracting the mean and then dividing by the standard deviation. We then applied the same to the test set using the sample mean and standard deviation from the training set. \\ \\
    
    \noindent \textbf{Models} \quad
    We applied 5 models in total. The baseline model was simply the average of all the census tracts' mean reported travel time. Because we believe the reported travel time is correlated with the grid time, we then fitted 3 linear regressions models: the first model used minimum grid time as the sole independent variable; the second model used average grid time as the sole independent variable; the third model included minimum grid time and average grid time as its independent variables. The last model is a LASSO (Least Absolute Shrinkage and Selection Operator) regression model incorporating all 19 socio-economic factors besides minimum grid time and average grid time. \\ \\ 
    
    \noindent \textbf{Average MSE on the test set} \quad
    To compare these 5 models, we calculated the mean squared error (MSE) on the test set for each fold. Then we took the average among these 5 folds. If the average MSE of the LASSO model incorporating socio-economic factors is lower than all the other models, then we have evidence to believe that socio-economic factors contribute to differences in reported travel time among various census tracts. Otherwise, we believe that the reported travel time solely depends on geo-location factors reflected by the grid time and is not explicitly affected by socio-economic factors.
    
    As reported in table \ref{tab:socio.eco}, we found that the lowest average MSE was achieved by the linear regression model fitted on the average grid time. Thus, we conclude that socio-economic factors may not explicitly affect the travel time for ambulances in Austin. 
    
\begin{table}[ht]
    \centering
    \begin{tabular}{|c|c|c|}\hline
    Model & Variable(s) & Average MSE\\\hline\hline
    Mean in the train set & N/A & 9.7980 \\
    Linear Regression & min.station.time & 9.6429 \\
    Linear Regression & avg.station.time & 9.5944 \\
    Linear Regression & min.station.time + avg.station.time & 9.6126 \\
    Lasso & All 21 variables & 9.7782 \\
    \hline
    \end{tabular}
    \caption{Comparing the 5 regression models.}
    \label{tab:socio.eco}
\end{table}

\section{User Guide}\label{sec:UserGuide}

\begin{figure}[ht]
    \centering
    \includegraphics[width=0.48\textwidth]{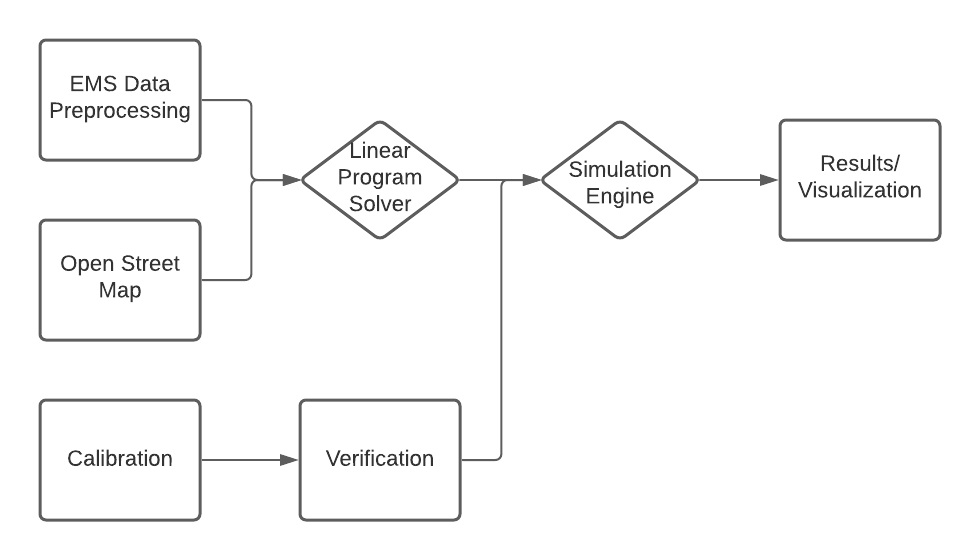}
    \caption{Main modules of the openEMS software package. Python modules are in squares, Julia modules are in diamonds.}
    \label{fig:software_state_space}
\end{figure}

OpenEMS is open-source and published at \url{https://github.com/joshua-ong/AmbulanceDeployment}. A user guide of the software package can be seen in figure \ref{fig:software_state_space}; an overview of the software variables are in table \ref{tab:variables}. 

\begin{tabularx}{\textwidth}{|>{\bfseries}lX|}
\hline
Variable & Description \\\hline\hline
$|$ regions $|$ & The number or regions is how many uniform discrete regions the continuous space is made into. This can be though of as the resolution of the grid.\\\hline
grid & An object containing the information of the discrete grid. This includes a time matrix between any two grid cells. \\\hline
adjacent neighborhood & Boolean matrix in $\{0, 1\}^{|regions|\times|regions|}$. Where index $i,j$ is 1 if region $i$ \newline is adjacent to region $j$.\\\hline
coverage  & Boolean Matrix in $\{0, 1\}^{|stations|\times|regions|}$ where index $i,j$ is 1 if station $i$ is within 10 minutes of region $j$. \\\hline
LP train calls  & Matrix where each row contains how many calls happened each hour. So entry $i,j$ is the number of calls at hour $i$ and region $j$. \\\hline
callibration train calls  &  Vector of train calls including time and location. \\\hline
test calls & Vector of test calls including time and location. \\\hline
callibration & Object that takes in grid time and outputs the adjusted grid time. \\\hline
lognormal$\_$params & This includes ($\mu$,$\sigma$) for a log normal distribution that are fit to the service time \\\hline 
$\textbf{x}^*$ & Integer vector in $\mathbb{Z}^{|Stations|}$ where index $i$ is the optimal number of ambulances in station $i$. \\
\hline
ncalls & number of calls to simulate  \\
\hline
nbatches & number of batches to simulate  \\
\hline
\caption{List of variables for user guide.}\label{tab:variables}\\
\end{tabularx} 

\textbf{Open Street Map} module creates the discrete grid described in section \ref{sec:gridConstruction}. Inputs include the number of regions, as well as the minimum/maximum longitudes/latitudes that bound the city of interest. The module returns the grid object which can be saved as a .json file. This grid object includes both the data about the grid cells as well as useful functions to handle that data. For socioeconomic interest, it can also contain a mapping of census tracts to grid points.  \\

$
\begin{array}{ll@{}ll}
> \texttt{grid = create$\_$grid(num\_regions,min$\_$lattitude,max$\_$lattitude} \\  
\texttt{,min$\_$longitude,max$\_$longitude) }  
\end{array} 
$\\

\textbf{Data Pre-processing} module collects the city EMS call data with headers for date of call, latitude and longitude of the emergency call for training the linear program and simulations. Further headers for reported travel time, reported service time, and latitude and longitude of the ambulance at the time of the call are needed for calibration and verification. The data is then processed into an adjacent neighborhood matrix, coverage matrix, LP train call matrix, calibration train call matrix, and test call matrix and saved. The user may consider customizing training and testing sets. For example, if the city's data changed drastically because of the Covid-19 pandemic, they could train before Covid-19 and test during Covid-19.   

$
\begin{array}{ll@{}ll}
> \texttt{(adjacent$\_$nbhd,coverage,LP$\_$train$\_$call,callibration$\_$train\_call,test\_call)}  \\
\texttt{=data\_pre$\_$processing(city$\_$data = "path/to/city$\_$data.csv",grid) }  
\end{array} 
$\\

\textbf{Calibration} module includes fitting a regression between grid times and reported travel times to calibrate the model. There is an example in section \ref{sec: case_study_callibration}. Datasets should be trained with the \texttt{callibration\_train\_set}. We included linear and log-log linear regressions, but we suggest the reader look at other regressions to find what works best for their dataset. The result is the adjusted travel time.

\textbf{Verification} module measures the accuracy of the adjusted travel time. There is an example in \ref{sec: case_study_verification}. The test calls vector includes the location and time of the call, as well as the reported response time and longitude and latitude of the ambulance at the time of the call. This is so we can verify the ambulance reported travel path time with our grid. The calibration object takes grid time and outputs adjusted grid time. The outputs are the differences in adjusted grid time and reported time. This can be measured as a score or as a plot.

$
\begin{array}{ll@{}ll}
> \texttt{verification(test\_call,callibration = log$\_$callibration)}  
\end{array} 
$\\

\textbf{Linear Program Solvers} module solves the two stage stochastic and robust programs detailed in section \ref{sec:Optimization}. Besides the pre-processed data, the other input is the $\alpha$ value that determines how conservative the robust model is. The function then returns the solved optimal stationing $x_{stoch^*}$ and $x_{robust^*}$.

$
\begin{array}{ll@{}ll}
>  x_{stoch^*},x_{robust^*} = \texttt{generate$\_$models($\alpha$ = .1,LPtrain$\_$calls,adjacent$\_$nbhd,} \\
\texttt{coverage,callibration=log$\_$callibration,grid)}  \\
\end{array}
$\\

\textbf{Simulation} module performs numerical computations using the test calls, optimal stationing, and adjusted grid time. There is an example in \ref{sec: case_study_results}. The inputs were all computed in previous sections. The output is the computed response time. From this one can compute their metrics.  

$
\begin{array}{ll@{}ll}
> \texttt{simulated\_response\_time = simulation(x,test$\_$calls,grid,n$\_$calls} \\
\texttt{= 1000,lognormal\_params = ($\mu$,$\sigma$),callibration = log$\_$callibration)}  
\end{array} 
$\\

\textbf{Results} The results will require unique analysis for each case study. All of the visualizations from this paper are included and can be reproduced. 

\section{Conclusion}\label{sec:Conclusion}

This paper presents an open-source project OpenEMS, an end-to-end pipeline to optimize ambulance strategic decisions. It includes two-stage stochastic and robust programming to optimize ambulance stationing and routing. In this open-source package, we include many of the details that prevent state of the art research from being applied on a case study level. This includes constructing a realistic grid using open street maps, calibrating these travel times to account for ambulances' special privileges, and verifying they work how we think they should work. Further, we provide practical guidance to hypothetical questions like the case where an EMS department has to add or remove one or more ambulances from their fleet. This is a common question.

We then applied this software to a case study on the city of Austin, Texas. In it, we found we could increase the mean response time by 88.02 seconds if the stationing was changed. We also showed the stochastic optimization performed better than the robust optimization. We also found the ideal locations to add or remove an ambulance to increase or maintain performance respectively. 

There is a lot of avenues for future work. Now that this code is open-source, one could simply try some new mathematical program and simulate for new results, for example, the distributionally robust optimization (DRO) that is growing in popularity in the operation research community. Along the lines of the simulation, we want to include more features such as sensitivity analysis with the choice of the grid resolution. There are also more questions about robust optimization; one main problem is the non-convex nature of the VaR constraint in the formulation. This leads to some unexpected results. For example, alpha = .1 and alpha = .01 could work fine, but this does not give any guarantees on interpolations between those values.

\section*{Funding details}

This research is supported by the Good Systems Research Initiative, part of University of Texas at Austin Bridging Barriers. Ngoc Mai Tran is supported by NSF Grant DMS-2113468 and the NSF IFML 2019844 award to the University of Texas at Austin. Evdokia Nikolova is supported by NSF CCF 1733832. \\

We also want to thank the Senior Project team from University of Texas at Austin who helped code sections of this open-source project: Ethan Santoni-Colvin, Michael Hilborn, Sudeep Narala, Xander Tedjo, Will Worthington

\section*{Disclosure statement}
The authors report there are no competing interests to declare. 

\section{References}

\bibliographystyle{plain}
\bibliography{references}

\begin{thebibliography}{10}

\bibitem{CDCSVI2018}
Centers for disease control and prevention/ agency for toxic substances and
  disease registry/ geospatial research, analysis, and services program.
\newblock CDC/ATSDR Social Vulnerability Index Insert 2018 Database Texas,
  Accessed on 08/08/2021.

\bibitem{ABOUELJINANE_2013}
L.~Aboueljinane, E.~Sahin, and Z.~Jemai.
\newblock A review on simulation models applied to emergency medical service
  operations.
\newblock {\em Computers and Industrial Engineering}, 66(4):734--750, 2013.

\bibitem{Aboueljinane2012R}
Lina Aboueljinane, Zied Jema{\"i}, and E.~Sahin.
\newblock Reducing ambulance response time using simulation: The case of
  val-de-marne department emergency medical service.
\newblock {\em Proceedings Title: Proceedings of the 2012 Winter Simulation
  Conference (WSC)}, pages 1--12, 2012.

\bibitem{ahmed}
Shabbir Ahmed.
\newblock {\em Two-Stage Stochastic Integer Programming: A Brief Introduction}.
\newblock American Cancer Society, 2011.

\bibitem{Batta1989}
Rajan Batta, June~M. Dolan, and Nirup~N. Krishnamurthy.
\newblock The maximal expected covering location problem: Revisited.
\newblock {\em Transportation Science}, 23(4):277--287, 1989.

\bibitem{BERALDI2004183}
P.~Beraldi, M.E. Bruni, and D.~Conforti.
\newblock Designing robust emergency medical service via stochastic
  programming.
\newblock {\em European Journal of Operational Research}, 158(1):183--193,
  2004.

\bibitem{bertsimas2019robust}
Dimitris Bertsimas and Yeesian Ng.
\newblock Robust and stochastic formulations for ambulance deployment and
  dispatch.
\newblock {\em European Journal of Operational Research}, 279(2):557--571,
  2019.

\bibitem{bezanson2015julia}
Jeff Bezanson, Alan Edelman, Stefan Karpinski, and Viral~B. Shah.
\newblock Julia: A fresh approach to numerical computing, 2015.

\bibitem{Birge_2011}
John~R. Birge and Fran{\c{c}}ois Louveaux.
\newblock {\em Introduction to Stochastic Programming}.
\newblock Springer New York, 2011.

\bibitem{BROWN2000}
Lawrence~H. Brown, Christa~L. Whitney, Richard~C. Hunt, Michael Addario, and
  Troy Hogue.
\newblock Do warning lights and sirens reduce ambulance response times?
\newblock {\em Prehospital Emergency Care}, 4(1):70--74, 2000.

\bibitem{BURKEY_2012}
M.L. Burkey, J.~Bhadury, and H.A. Eiselt.
\newblock A location-based comparison of health care services in four u.s.
  states with efficiency and equity.
\newblock {\em Socio-Economic Planning Sciences}, 46(2):157--163, 2012.
\newblock Modeling Public Sector Facility Location Problems.

\bibitem{Courtemanche2019Sep}
Charles Courtemanche, Andrew Friedson, Andrew~P. Koller, and Daniel~I. Rees.
\newblock {The affordable care act and ambulance response times}.
\newblock {\em J. Health Econ.}, 67:102213, Sep 2019.

\bibitem{Doerner}
Karl~F. Doerner and Richard~F. Hartl.
\newblock Health care logistics, emergency preparedness, and disaster relief:
  New challenges for routing problems with a focus on the austrian situation.
\newblock In {\em Operations Research/Computer Science Interfaces}, pages
  527--550. Springer {US}, 2008.

\bibitem{Eaton_1985}
David~J. Eaton, Mark~S. Daskin, Dennis Simmons, Bill Bulloch, and Glen Jansma.
\newblock Determining emergency medical service vehicle deployment in austin,
  texas.
\newblock {\em Interfaces}, 15(1):96--108, feb 1985.

\bibitem{Fitzgerald2005Sep}
Guy Fitzgerald and Nancy~L. Russo.
\newblock {The turnaround of the London Ambulance Service Computer-Aided
  Despatch system (LASCAD)}.
\newblock {\em European Journal of Information Systems}, 14(3):244--257, Sep
  2005.

\bibitem{GENDREAU199775}
Michel Gendreau, Gilbert Laporte, and Frédéric Semet.
\newblock Solving an ambulance location model by tabu search.
\newblock {\em Location Science}, 5(2):75--88, 1997.

\bibitem{Haklay2008OpenStreetMapUS}
M.~Haklay and Patrick Weber.
\newblock Openstreetmap: User-generated street maps.
\newblock {\em IEEE Pervasive Computing}, 7:12--18, 2008.

\bibitem{Markovitch1995}
J.E. Hammann and N.A. Markovitch.
\newblock Introduction to arena [simulation software].
\newblock In {\em Winter Simulation Conference Proceedings, 1995.}, pages
  519--523, 1995.

\bibitem{Price2000}
C.R. Harrell and R.N. Price.
\newblock Simulation modeling and optimization using promodel.
\newblock In {\em 2000 Winter Simulation Conference Proceedings (Cat.
  No.00CH37165)}, volume~1, pages 197--202 vol.1, 2000.

\bibitem{Heightman2009Mar}
A.~J. Heightman.
\newblock {Resource overkill: we can do more for less}.
\newblock {\em JEMS}, 34(3):16., Mar 2009.

\bibitem{Henderson2005}
Shane~G. Henderson and Andrew~J. Mason.
\newblock {Ambulance Service Planning: Simulation and Data Visualisation}.
\newblock In {\em {Operations Research and Health Care}}, pages 77--102.
  Springer, Boston, MA, Boston, MA, USA, 2005.

\bibitem{Ingolfsson_2003}
A~Ingolfsson, E~Erkut, and S~Budge.
\newblock Simulation of single start station for edmonton ems.
\newblock {\em Journal of the Operational Research Society}, 54(7):736--746,
  2003.

\bibitem{Jagtenberg_2017Dec}
C.~J. Jagtenberg, S.~Bhulai, and R.~D. van~der Mei.
\newblock {Dynamic ambulance dispatching: is the closest-idle policy always
  optimal?}
\newblock {\em Health Care Manag. Sci.}, 20(4):517--531, Dec 2017.

\bibitem{Inakawa_2010}
Takehiro Furuta Atsuo~Suzuki Keisuke~Inakawa.
\newblock Effect of ambulance station locations and number of ambulances to the
  quality of the emergency service.
\newblock In {\em Ninth International Symposium on Operations Research and Its
  Applications (ISORA’10) Chengdu-Jiuzhaigou}, Jun 2010.
\newblock [Online; accessed 19. Jun. 2021].

\bibitem{Lam_2016}
Sean Shao~Wei Lam, Yee~Sian Ng, Mohanavalli~Rajagopal Lakshmanan, Yih~Yng Ng,
  and Ong. Marcus~Eng Hock.
\newblock Ambulance deployment under demand uncertainty.
\newblock {\em Journal of Advanced Management Science}, 2016.

\bibitem{Lam2016}
Sean Shao~Wei Lam, Yee~Sian Ng, Mohanavalli~Rajagopal Lakshmanan, Yih~Yng Ng,
  and Marcus Eng~Hock Ong.
\newblock Ambulance deployment under demand uncertainty.
\newblock {\em Journal of Advanced Management Science Vol}, 4(3), 2016.

\bibitem{LIU201979}
Kanglin Liu, Qiaofeng Li, and Zhi-Hai Zhang.
\newblock Distributionally robust optimization of an emergency medical service
  station location and sizing problem with joint chance constraints.
\newblock {\em Transportation Research Part B: Methodological}, 119:79--101,
  2019.

\bibitem{lubin_2015}
Miles Lubin and Iain Dunning.
\newblock {Computing in Operations Research Using Julia}.
\newblock {\em INFORMS Journal on Computing}, 27(2):238--248, May 2015.

\bibitem{Mahama2018Dec}
Mohammed-Najeeb Mahama, Ernest Kenu, Delia~Akosua Bandoh, and Ahmed~Nuhu
  Zakariah.
\newblock {Emergency response time and pre-hospital trauma survival rate of the
  national ambulance service, Greater Accra (January {\textendash} December
  2014)}.
\newblock {\em BMC Emerg. Med.}, 18(1):1--7, Dec 2018.

\bibitem{Mason_2013}
Andrew~James Mason.
\newblock Simulation and real-time optimised relocation for improving ambulance
  operations.
\newblock In {\em Handbook of Healthcare Operations Management}, pages
  289--317. Springer New York, 2013.

\bibitem{Maxwell_2010}
Matthew~S. Maxwell, Mateo Restrepo, Shane~G. Henderson, and Huseyin Topaloglu.
\newblock Approximate dynamic programming for ambulance redeployment.
\newblock {\em {INFORMS} Journal on Computing}, 22(2):266--281, may 2010.

\bibitem{mccormack_2015}
R.J. McCormack and G.~Coates.
\newblock A simulation model to enable the optimization of ambulance fleet
  allocation and base station location for increased patient survival.
\newblock {\em European journal of operational research.}, 247(1):294--309, May
  2015.

\bibitem{McLay_2013}
Laura~A. McLay and Maria~E. Mayorga.
\newblock A model for optimally dispatching ambulances to emergency calls with
  classification errors in patient priorities.
\newblock {\em {IIE} Transactions}, 45(1):1--24, jan 2013.

\bibitem{Morohosi2008ACS}
Hozumi Morohosi.
\newblock A case study of optimal ambulance location problems.
\newblock {\em Operations Research and its Applications (The Sixth
  International Symposium, ISORA’08 Proceedings)}, 2008.

\bibitem{Nilsang_2018}
Suriyaphong Nilsang, Chumpol Yuangyai, Chen-Yang Cheng, and Udom Janjarassuk.
\newblock Locating an ambulance base by using social media: a case study in
  bangkok.
\newblock {\em Annals of Operations Research}, 283(1-2):497--516, jun 2018.

\bibitem{Pons2002Jul}
Peter~T. Pons and Vincent~J. Markovchick.
\newblock {Eight minutes or less: does the ambulance response time guideline
  impact trauma patient outcome?1 1Selected Topics: Prehospital Care is
  coordinated by Peter T. Pons, MD, of Denver Health Medical Center, Denver,
  Colorado}.
\newblock {\em J. Emerg. Med.}, 23(1):43--48, Jul 2002.

\bibitem{ReVelle_1989}
Charles ReVelle and Kathleen Hogan.
\newblock The maximum availability location problem.
\newblock {\em Transportation Science}, 23(3):192--200, aug 1989.

\bibitem{SCHMID2012}
Verena Schmid.
\newblock Solving the dynamic ambulance relocation and dispatching problem
  using approximate dynamic programming.
\newblock {\em European Journal of Operational Research}, 219(3):611--621,
  2012.
\newblock Feature Clusters.

\bibitem{SWALEHE2016199}
Masoud Swalehe and Semra~Gunay Aktas.
\newblock Dynamic ambulance deployment to reduce ambulance response times using
  geographic information systems: A case study of odunpazari district of
  eskisehir province, turkey.
\newblock {\em Procedia Environmental Sciences}, 36:199--206, 2016.
\newblock International Conference on Geographies of Health and Living in
  Cities: Making Cities Healthy for All.

\bibitem{tassone2020comprehensive}
Joseph Tassone and Salimur Choudhury.
\newblock A comprehensive survey on the ambulance routing and location
  problems, 2020.

\bibitem{White_2007}
Chapin White.
\newblock Health care spending growth: How different is the united states from
  the rest of the {OECD}?
\newblock {\em Health Affairs}, 26(1):154--161, jan 2007.

\bibitem{ZENG2013457}
Bo~Zeng and Long Zhao.
\newblock Solving two-stage robust optimization problems using a
  column-and-constraint generation method.
\newblock {\em Operations Research Letters}, 41(5):457--461, 2013.

\end{thebibliography}

\appendix

\section{Additional figures for numerical results} \label{app:numerical_appendix}

This section presents additional tables and figures for the numerical experiments.

\begin{figure}[ht]
    \centering
    \includegraphics[width=0.48\textwidth]{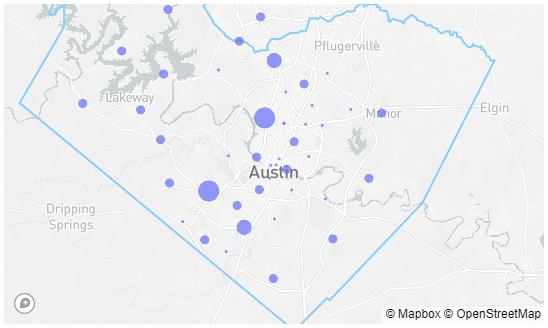}
    \includegraphics[width=0.48\textwidth]{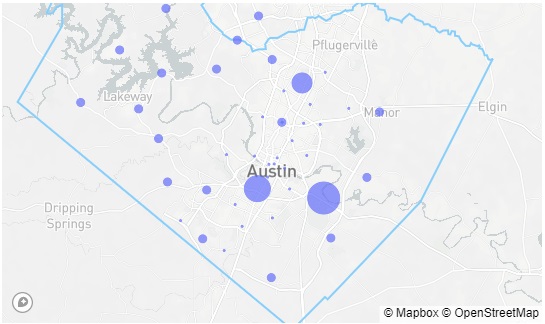}
    \includegraphics[width=0.48\textwidth]{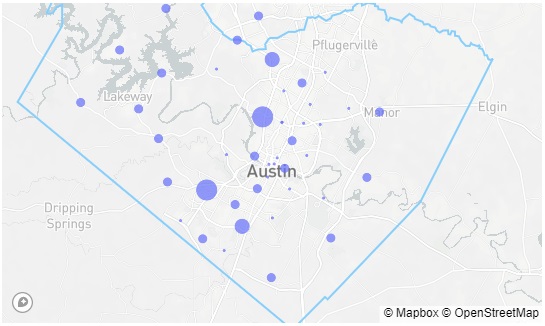}
    \includegraphics[width=0.48\textwidth]{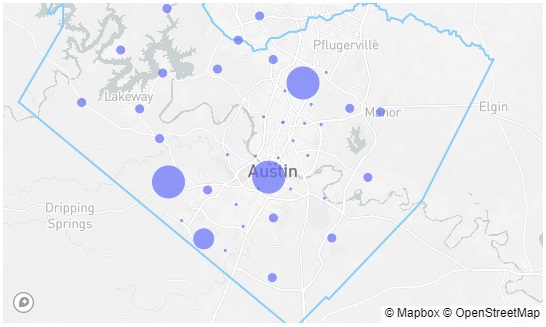}
    \includegraphics[width=0.48\textwidth]{stochastic40.jpeg}
    \includegraphics[width=0.48\textwidth]{robust40.jpeg}
    \includegraphics[width=0.48\textwidth]{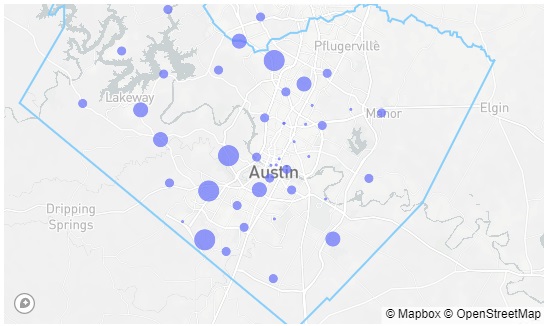}
    \includegraphics[width=0.48\textwidth]{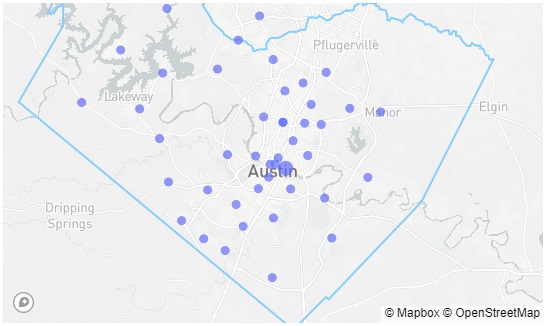}
    \includegraphics[width=0.48\textwidth]{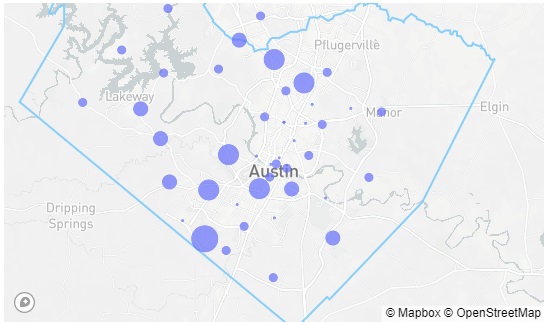}
    \includegraphics[width=0.48\textwidth]{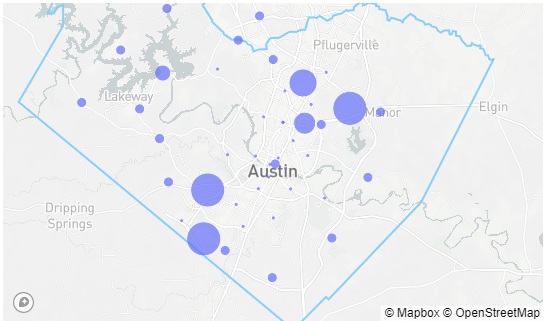}
    \caption{Ambulance location schemes for the stochastic optimization (left column) and robust ($\alpha$ = .01) optimization (right column). Rows start with 30 ambulances, increase by 5 ambulances per row, and end with 50 ambulances.
    The purple circle indicates there is an ambulance station there. 
    The radius of the circle is proportional to the number of ambulances in the area where the biggest circle is 5 and the smallest circle is 0.}
    \label{fig:all_amb_stationing}
\end{figure}

\begin{figure}[ht]
\centering
\includegraphics[width=1.0 \textwidth]{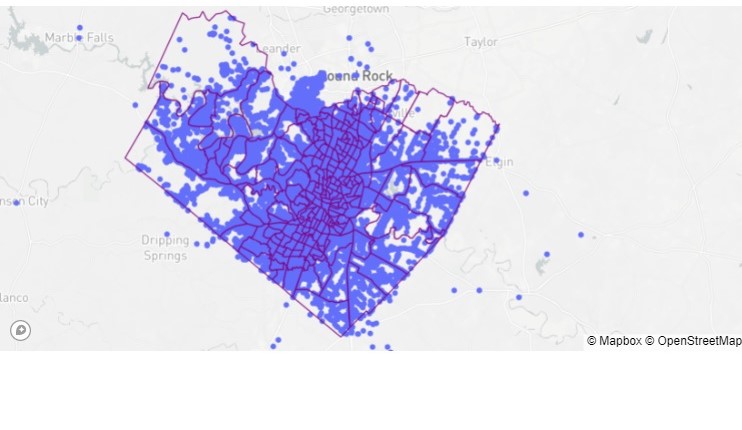}
\caption{A scatter plot of all call data. Note they cannot represent density.}
\label{fig:geo_scatter}
\end{figure}

\begin{figure}[ht]
\centering
\includegraphics[width=1.0 \textwidth]{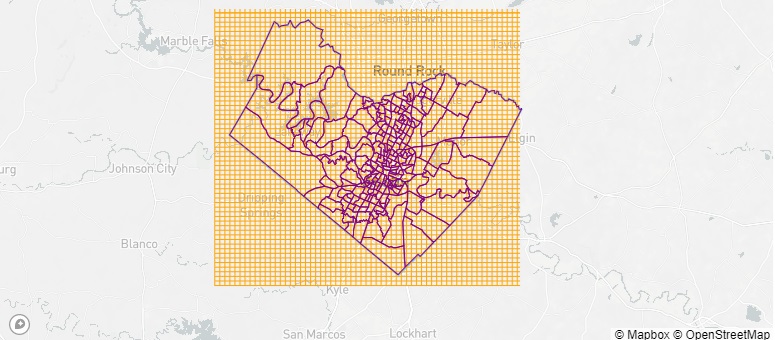}
\caption{A visualization of the discrete grid over Austin county.}
\label{fig:grid_plot}
\end{figure}

\begin{table}[ht]
    \centering
    \begin{tabular}{|l|l|l|l|l|l|}\hline
     & $\alpha$ = .1 &  $\alpha$= .05 &   $\alpha$ = .01 &   $\alpha$ = .001 &   $\alpha$ = .0001    \\\hline\hline
       cross validation 1 & 6.807&   6.562 &    6.562 &    6.562 &   6.562  \\\hline
       cross validation 2 &  6.806 &  6.562 &    6.562 &    6.562 &    7.190  \\\hline
       cross validation 3 & 6.562 &  6.562  &    6.562 &    6.562 &   7.590   \\\hline
    \end{tabular}
    \caption{Three-fold crossvalidation on 5 alpha values to find when $\alpha$ saturates.}
    \label{tab:cross_validation}
\end{table}

\section{List of variables extracted from CDC SVI dataset}

\begin{table}[ht]
    \centering
    \begin{tabular}{|l|l|}\hline
    Variable & Description \\\hline\hline
        E\_TOTPOP & Population estimate\\\hline
        E\_HU & Housing units estimate\\\hline
        E\_HH & Households estimate\\\hline
        E\_POV & Persons below poverty estimate\\\hline
        E\_UNEMP & Civilian (age 16+) unemployed estimate\\\hline
        E\_NOHSDP & Persons (age 25+) with no high school diploma estimate \\\hline
        E\_AGE65 & Persons aged 65 and older estimate \\\hline
        E\_AGE17 & Persons aged 17 and younger estimate \\\hline
        E\_DISABL& Civilian noninstitutionalized population with a disability estimate\\\hline
        E\_SNGPNT & Single parent household with children under 18 estimate \\\hline
        E\_MINRTY & Minority (all persons except white, non-Hispanic) estimate \\\hline
        E\_LIMENG & Persons (age 5+) who speak English "less than well" estimate \\\hline
        E\_MUNIT & Housing in structures with 10 or more units estimate \\\hline
        E\_MOBILE & Mobile homes estimate \\\hline
        E\_CROWD & At household level (occupied housing units), more people than rooms estimate \\\hline
        E\_NOVEH & Households with no vehicle available estimate \\\hline
        E\_GROUPQ & Persons in group quarters estimate \\\hline
        E\_UNINSUR & Uninsured in the total civilian noninstitutionalized population estimate \\\hline
        E\_DAYPOP & Estimated daytime population \\\hline
    \end{tabular}
    \caption{List of variables extracted from CDC SVI dataset}
    \label{tab:SVI list}
\end{table}

\end{document}